\documentclass[11pt,a4paper]{article}
\usepackage{jheppub}
\usepackage{graphicx, color}
\usepackage{stmaryrd}
\usepackage{amsmath, amssymb}

\newcommand*{\nc}{\newcommand*}
\nc{\real}{\mathbb{R}}
\nc{\comp}{\mathbb{C}}
\nc{\nat}{\mathbb{N}}
\nc{\integ}{\mathbb{Z}}
\nc{\mc}{\mathcal}
\nc{\ca}{{\mathcal A}}
\nc{\cd}{{\mathcal D}}
\nc{\dee}{{\mathcal D}}
\nc{\ih}{{\mathcal I}}
\nc{\lag}{{\mathcal L}}
\nc{\cm}{{\mathcal M}}
\nc{\oh}{{\mathcal O}}
\nc{\cq}{{\mathcal Q}}
\nc{\sig}{\sigma}
\nc{\gam}{\gamma}
\nc{\kap}{\kappa}
\nc{\lam}{\lambda}
\nc{\Gam}{\Gamma}
\nc{\Lam}{\Lambda}
\nc{\eps}{\varepsilon}
\nc{\be}{\begin{equation}}
\nc{\ee}{\end{equation}}
\nc{\ben}{\begin{equation*}}
\nc{\een}{\end{equation*}}
\nc{\ba}{\begin{aligned}}
\nc{\ea}{\end{aligned}}
\nc{\barr}{\left( \begin{array}}
\nc{\earr}{\end{array} \right)}
\nc{\new}{\\[5 mm]}
\nc{\tab}{\hspace{5 mm}}

\renewcommand{\it}{\textit}
\renewcommand{\bf}{\textbf}
\nc{\half}{\frac{1}{2}}
\nc{\Tr}{\text{Tr}}
\nc{\der}{\partial}
\nc{\derb}{\bar{\partial}}
\nc{\p}{\partial}
\nc{\xh}{\hat{x}}
\nc{\yh}{\hat{y}}
\nc{\zb}{{\bar{z}}}
\nc{\bz}{\zb}
\nc{\wb}{{\bar{w}}}
\nc{\para}{\parallel}
\nc{\lra}{\leftrightarrow}
\nc{\llra}{\longleftrightarrow}
\nc{\lang}{\langle}
\nc{\rang}{\rangle}
\nc{\ket}[1]{{\mid #1 \, \rangle}}
\nc{\bra}[1]{{\langle \, #1 \mid}}
\nc{\inner}[2]{\langle \, #1 \mid #2 \, \rangle}
\nc{\al}{\alpha'}
\nc{\kket}[1]{{\mid #1 \, \rang \rang}}
\nc{\bbra}[1]{{\lang \lang \, #1 \mid}}
\nc{\req}{\stackrel{\real}{=}}
\nc{\ceq}{\stackrel{\comp}{=}}
\newcommand{\res}[1]{\underset{#1}{\rm Res}}

\begin{document}

\title{Currents in Celestial CFT}

\author{Adam Ball}

\affiliation{Perimeter Institute for Theoretical Physics, 31 Caroline St\\
Waterloo, ON N2L 2Y5,
Canada\\
aball1@pitp.ca}

\abstract{In this review we discuss currents in celestial CFT and the consistency of their na\"ive symmetry algebras. In particular we study in detail the Jacobi identity and the double residue condition for soft insertions, hard momentum space insertions, and hard celestial insertions. In the latter case we introduce the notion of a ``hard current" in CFT and work through examples in the 2D critical Ising model. The current algebra of hard insertions in pure Einstein gravity is a slight conceptual generalization of the familiar $w_{1+\infty}$-wedge current algebra. We also review branch cut terms in the celestial OPE, which indicate new primary content and were previously missed until recently. We work through an explicit toy example illustrating the mechanism by which such branch cut terms can arise. These branch cut terms prevent a symmetry interpretation but are fully compatible with a consistent OPE.}

\maketitle

\section{Introduction}

Celestial holography recasts scattering amplitudes in (asymptotically) flat spacetime as conformal correlators in a putative codimension-two theory. The general approach has been expertly reviewed elsewhere \cite{Raclariu:2021zjz, Pasterski:2021rjz, Pasterski:2021raf, Pasterski:2023ikd}. The program grew out of the observation that Weinberg's soft theorems \cite{Weinberg:1965nx} can be interpreted as the Ward identities of asymptotic symmetries \cite{He:2014laa, He:2014cra, He:2015zea}, and its guiding principle is to put symmetries front and center. This has led to the discovery of infinite-dimensional symmetry algebras in the tree-level $\mc{S}$-matrices of Einstein gravity and Yang-Mills theory \cite{Guevara:2021abz, Strominger:2021mtt}, as well as in certain theories at loop level. The construction of the infinite-dimensional symmetry algebras in Ref. \cite{Guevara:2021abz} has a straightforward na\"ive generalization to any theory at tree-level, but it was shown in Refs. \cite{Mago:2021wje, Ren:2022sws} that it generically runs into consistency problems. Specifically, the na\"ive commutator of charges fails the Jacobi identity. In this review we explain why this happens and argue that it poses no problem for associativity of the OPE.

Most work on celestial holography has been at tree level, and we will continue this trend here. Our interest is primarily in soft and collinear poles, so we will also restrict attention to massless theories for convenience. We work in four spacetime dimensions throughout.

\section{Celestial basics}

The $\mc{S}$-matrix is traditionally formulated in terms of plane waves, which are translation eigenstates. In celestial holography we opt instead for boost eigenstates. This change of basis can be implemented directly at the level of the $\mc{S}$-matrix by Mellin transforming the (lightcone) energy of each particle. We parametrize massless momenta as
\be p^\mu(\omega, z, \zb; \epsilon) = \epsilon \, \omega (1 + z\zb, z + \zb, -i(z-\zb), 1 - z\zb), \ee
where $\epsilon=\pm 1$ determines whether the particle is incoming or outgoing in the ``all-outgoing" convention. A celestial amplitude is then defined by
\be \label{eq:celamp} \tilde\ca_n(\Delta_1, z_1, \zb_1; \dots ;\Delta_n, z_n, \zb_n) \equiv \int_0^\infty d\omega_1 \omega_1^{\Delta_1-1} \dots \int_0^\infty d\omega_n \omega_n^{\Delta_n-1} \ca_n(p_1, \dots, p_n) \ee
where $\ca_n(p_1, \dots, p_n)$ is an \it{unstripped} $n$-point scattering amplitude, i.e. it includes the momentum-conserving delta function. We will write stripped amplitudes with a hat:
\be \ca_n(p_1, \dots, p_n) = \delta^{(4)}\big(\sum_{i=1}^n p_i\big) \, \hat\ca_n(p_1, \dots, p_n). \ee
We will often suppress discrete labels like $\epsilon_i$. Under the natural isomorphism of the 4D Lorentz group with the 2D conformal group, celestial amplitudes transform precisely as 2D conformal correlators, with the $\Delta_i$ as the conformal dimensions. To emphasize this point of view we often adopt the suggestive notation
\be \lang \oh_{\Delta_1}(z_1, \zb_1) \dots \oh_{\Delta_n}(z_n, \zb_n) \rang \equiv \tilde\ca_n(\Delta_1, z_1, \zb_1; \dots ;\Delta_n, z_n, \zb_n). \ee
We note here that the 2D spin of the operator $\oh_\Delta$ coincides with the 4D helicity of the particle it is constructed from. It is often convenient to use a similar notation in momentum space. Accordingly we define
\be \lang \oh(\omega_1, z_1, \zb_1) \dots \oh(\omega_n, z_n, \zb_n) \rang \equiv \ca_n(p_1, \dots, p_n). \ee
The allowed range of $\Delta_i$ is still under debate. Initially Ref. \cite{Pasterski:2017kqt} argued in favor of the principal series $\Delta \in 1 + i\real$, but gradually more authors \cite{Atanasov:2021oyu, Adamo:2021lrv, Freidel:2022skz, Cotler:2023qwh} are exploring the discrete values $\Delta \in \integ$. See Ref. \cite{Mitra:2024ugt} for a recent discussion. Brushing aside issues of completeness, at the level of analytic continuation one can ask about any complex value of $\Delta$, and this is the rather agnostic stance we will take. Another subtle point is that we work with EFTs valid only at low energy, yet the Mellin transform involves energies of all scales \cite{Arkani-Hamed:2020gyp}. As explained later, conformally soft limits probe low energy information about the original amplitude, so this will mostly not be a problem for us.

Since the energy $\omega$ is not a parameter of celestial amplitudes, they cannot obey traditional soft theorems. Instead they obey the closely related \it{conformally} soft theorems. When the dimension $\Delta$ of a particle approaches an integer $k \le 1$, a pole like $\frac{1}{\Delta - k}$ develops. The residue of the pole contains certain universal information about the celestial amplitude with the conformally soft particle omitted. The first case studied was that of gluons with $\Delta \to 1$ in pure Yang-Mills \cite{Fan:2019emx, Pate:2019mfs, Nandan:2019jas}. Write $\oh^{a,s}_\Delta(z, \zb)$ for an outgoing gluon of helicity $s=\pm$ and dimension $\Delta$ with adjoint color index $a$. The conformally soft theorem for a positive-helicity gluon then reads
\be \ba \res{\Delta\to 1} \lang \oh^{a,+}_\Delta(z, \zb) & \oh^{a_1, s_1}_{\Delta_1}(z_1, \zb_1) \dots \oh^{a_n, s_n}_{\Delta_n}(z_n, \zb_n) \rang = \\
& \sum_{i=1}^n \frac{-i f^{a a_i}{}_b}{z - z_i} \lang \oh^{a_1, s_1}_{\Delta_1}(z_1, \zb_1) \dots \oh^{b,s_i}_{\Delta_i}(z_i, \zb_i) \dots \oh^{a_n, s_n}_{\Delta_n}(z_n, \zb_n) \rang \ea \ee
where the $f^{aa_i}{}_b$ are structure constants and summation over the repeated index $b$ is implied. The negative helicity version simply uses $\zb-\zb_i$ in place of $z-z_i$. This conformally soft factor is precisely the same as the soft factor in the usual momentum space soft gluon theorem. Note that all $\zb$ dependence has dropped out, and the $z$ dependence is rational, with poles only at the other insertions. As such, the soft gluon insertion can be interpreted as a holomorphic current in the celestial CFT, generating infinitely many symmetries. A similar story holds for gravitons \cite{Puhm:2019zbl}. It was later discovered in Ref. \cite{Guevara:2021abz} that there is an infinite tower of conformally soft theorems in gauge theory and gravity corresponding to the residue at $\Delta\to k$ for integers $k \le 1$, each of which gives rise to (generalized) holomorphic currents in celestial CFT. They can all be understood in terms of the following fact about the Mellin transform \cite{Guevara:2019ypd}. Let $g(\omega)$ be a function, and
\be \tilde g(\Delta) \equiv \int_0^\infty d\omega \, \omega^{\Delta-1} g(\omega) \ee
its Mellin transform. Defining the Laurent coefficients by\footnote{We assume $g(\omega)$ has only a simple pole at $\omega=0$ since we have amplitudes in mind.}
\be g(\omega) = \sum_{k=-\infty}^1 \frac{g_{(k)}}{\omega^k}, \ee
we have
\be \res{\Delta\to k} \, \tilde g(\Delta) = g_{(k)} \ee
for any integer $k \le 1$. Applying this to the $\omega$ dependence of an amplitude, we see that the conformally soft theorems are simply picking out the Laurent coefficients of the $\omega$ expansion back in momentum space. As such, there is little difference between energetically and conformally soft insertions.\footnote{There are caveats when we later consider taking multiple particles soft simultaneously. One should also keep in mind that this statement is about a particular particle; changing the basis for the \it{other}, hard, particles in the amplitude will of course change the amplitude.} This explains, for example, why the conformally soft factor for $\Delta\to 1$ is the same as the usual soft factor from the leading $1/\omega$ soft pole. It also provides a justification for discussing conformally soft theorems in the context of EFT, where the high energy behavior of the amplitudes is not even defined.

This momentum space perspective also gives us a bit more control over the global analytic properties of soft insertions. Recall that in momentum space tree-level amplitudes are rational functions of the momentum components, and in particular of the parameters $\omega_i, z_i, \zb_i$. The denominators of these rational functions come only from propagators, i.e. the squares of sums of subsets of the external momenta. Propagators involving the sum of only two external momenta have an important qualitative difference from those involving the sum of three or more external momenta. Setting $\epsilon_i = 1$ for simplicity, we have
\be \frac{1}{(p_1 \hspace{-.8mm} + \hspace{-.8mm} p_2)^2} = \frac{-1/4}{\omega_1\omega_2 z_{12} \zb_{12}}, \qquad \hspace{-1mm} \frac{1}{(p_1 \hspace{-.8mm} + \hspace{-.8mm} p_2 \hspace{-.8mm} + \hspace{-.8mm} p_3)^2} = \frac{-1/4}{\omega_1\omega_2 z_{12} \zb_{12} \hspace{-.8mm} + \hspace{-.8mm} \omega_1\omega_3 z_{13} \zb_{13} \hspace{-.8mm} + \hspace{-.8mm} \omega_2\omega_3 z_{23} \zb_{23}}. \ee
Both types of propagators give poles in, say, $z_1$. But as $\omega_1\to 0$, the $z_1$ pole from the three-particle (or higher-particle) propagator recedes to infinity. In contrast, for any value of $\omega_1$, the two-particle propagator gives a pole at $z_1 = z_2$. Since the Laurent expansion in $\omega_1$ is centered at $\omega_1=0$, its Laurent coefficients can only acquire $z_1$ poles from the two-particle propagators, i.e. the collinear singularities. A more precise version of this argument is given in Ref. \cite{Ball:2022bgg}. The upshot is that soft insertions can always be written in the general form
\be \label{eq:gensoft} \oh_{(k)}(z, \zb) \equiv \res{\Delta\to k} \oh_\Delta(z, \zb) = \sum_i \frac{f_i^{(k)}(\zb)}{z - z_i} + P^{(k)}(z, \zb) \ee
where the sum is over the other particles in the amplitude, $P^{(k)}(z, \zb)$ is a polynomial in $z$ (but not necessarily in $\zb$), and $f_i^{(k)}(\zb)$ is determined by the soft-collinear limit with particle $i$ or equivalently\footnote{The equivalence is up to a caveat, to be discussed in section \ref{sec:newOPE}.} the soft-hard celestial OPE, which is easily computed in practice. In celestial contexts the use of analytic continuation to treat $z, \zb$ as independent complex variables is standard. This is closely related to Wick rotating from Minkowski space to Klein space \cite{Atanasov:2021oyu}. With this in mind, the $z$ dependence of $\oh_{(k)}(z, \zb)$ has the properties of a holomorphic current: no branch cuts, and poles only at other insertions. Thus we can integrate it along a contour in $z$ (with $\zb$ fixed) to get a conserved charge, i.e. a mode. The nontrivial $\zb$ dependence makes it debatable whether $\oh_{(k)}(z, \zb)$ is a bona fide holomorphic current, but it does not obstruct the calculation of mode commutators nor the association with a symmetry algebra. However, as we will see, there can be other subtle obstructions to the symmetry interpretation.

\section{Currents and algebras}

For concreteness let us carry this out in pure Yang-Mills theory. The holomorphic collinear limits of an outgoing, positive-helicity gluon with another outgoing gluon are
\be \oh^{a,+}(\omega_1, z_1, \zb_1) \oh^{b,+}(\omega_2, z_2, \zb_2) \sim \frac{-if^{ab}{}_c}{z_{12}} \frac{\omega_1+\omega_2}{\omega_1\omega_2} \oh^{c,+}(\omega_1+\omega_2, z_2, \frac{\omega_1\zb_1+\omega_2\zb_2}{\omega_1+\omega_2}), \ee
\be \oh^{a,+}(\omega_1, z_1, \zb_1) \oh^{b,-}(\omega_2, z_2, \zb_2) \sim \frac{-if^{ab}{}_c}{z_{12}} \frac{\omega_2}{\omega_1 (\omega_1+\omega_2)} \oh^{c,-}(\omega_1+\omega_2, z_2, \frac{\omega_1\zb_1+\omega_2\zb_2}{\omega_1+\omega_2}), \ee
where $\sim$ indicates equality of $O(1/z_{12})$ pieces. We are treating $z, \zb$ as independent complex variables here. It was shown in Refs. \cite{Fan:2019emx, Pate:2019lpp} that Mellin transforming on $\omega_1, \omega_2$ leads to the celestial OPEs
\be \oh^{a,+}_{\Delta_1}(z_1, \zb_1) \oh^{b,+}_{\Delta_2}(z_2, \zb_2) \sim \frac{-i f^{ab}{}_c}{z_{12}} \sum_{m=0}^\infty B(\Delta_1-1+m, \Delta_2-1) \frac{\zb_{12}^m}{m!} \p_{\zb_2}^m \oh^{c,+}_{\Delta_1+\Delta_2-1}(z_2, \zb_2), \ee
\be \oh^{a,+}_{\Delta_1}(z_1, \zb_1) \oh^{b,-}_{\Delta_2}(z_2, \zb_2) \sim \frac{-i f^{ab}{}_c}{z_{12}} \sum_{m=0}^\infty B(\Delta_1-1+m, \Delta_2+1) \frac{\zb_{12}^m}{m!} \p_{\zb_2}^m \oh^{c,-}_{\Delta_1+\Delta_2-1}(z_2, \zb_2), \ee
where $B(x,y) = \frac{\Gam(x)\Gam(y)}{\Gam(x+y)}$ is the Euler beta function. Neither gluon has been taken soft yet, so these are hard-hard OPEs. Taking the first gluon soft gives the soft-hard OPEs
\be \oh^{a, +}_{(k)}(z_1, \zb_1) \oh^{b,+}_{\Delta_2}(z_2, \zb_2) \sim \frac{-i f^{ab}{}_c}{z_{12}} \sum_{m=0}^{1-k} {2-k-\Delta_2-m \choose 1-\Delta_2} \frac{\zb_{12}^m}{m!} \p_{\zb_2}^m \oh^{c,+}_{\Delta_2+k-1}(z_2, \zb_2), \ee
\be \oh^{a, +}_{(k)}(z_1, \zb_1) \oh^{b,-}_{\Delta_2}(z_2, \zb_2) \sim \frac{-i f^{ab}{}_c}{z_{12}} \sum_{m=0}^{1-k} {-k-\Delta_2-m \choose -1-\Delta_2} \frac{\zb_{12}^m}{m!} \p_{\zb_2}^m \oh^{c,-}_{\Delta_2+k-1}(z_2, \zb_2). \ee
There are similar formulas for one outgoing and one incoming gluon \cite{Pate:2019lpp}. We can use these soft-hard OPEs to read off the $f^{(k)}_i(\zb)$ functions in the general form for $\oh^{a,+}_{(k)}(z, \zb)$ above in \eqref{eq:gensoft}. They are polynomials in $\zb$ of degree $1-k$. The $z$-polynomial $P^{(k)}(z, \zb)$ is non-universal, and therefore more difficult to determine, but na\"ively one expects it not to affect the mode commutators since it contains no poles in $z$. Thus at the level of current algebras it is often irrelevant, and we will speak of ``holomorphic currents" that nonetheless have $z$-polynomial parts depending on $\zb$. One can construct such holomorphic currents by taking a soft gluon and mode expanding in $\zb$. This is the approach taken in Ref. \cite{Guevara:2021abz}, where they define the currents $R^{k,a}_n(z)$ by
\be \oh^{a,+}_{(k)}(z, \zb) = \sum_{n=\frac{k-1}{2}}^{n=\frac{1-k}{2}} \frac{R^{k,a}_n(z)}{\zb^{n+\frac{k-1}{2}}} \quad + \quad \text{polynomial in $z$}. \ee
One can isolate the currents using contour integrals in $\zb$. Alternatively, since the currents are the coefficients of a $\zb$ polynomial of degree $1-k$, one can project onto them (up to terms polynomial in $z$) using certain differential operators in $\zb$. For example $R^{k,a}_{\frac{k-1}{2}}(z)$, which is the coefficient of $\zb^{1-k}$, can simply be identified as $\frac{1}{(1-k)!} \p_\zb^{1-k} \oh^{a,+}_{(k)}(z, \zb)$. Note that this latter approach never requires choosing a value for $\zb$, nor a contour over which to integrate it. In practice it is convenient to leave the holomorphic currents packaged together in the single object $\oh^{a,+}_{(k)}(z, \zb)$, and we will usually do so.

With a current in hand, the natural next step is to compute commutators. Since the celestial OPE in Yang-Mills contains only simple poles in $z$, the commutators of modes of holomorphic currents are entirely encapsulated in the holomorphic commutator \cite{Guevara:2021abz} between the currents themselves,
\be [A, B](z) \equiv \oint_z \frac{dw}{2\pi i} A(w) B(z). \ee
This simply grabs the coefficient of $1/z_{12}$ in the OPE. However, right away we run into a problem. If we have two soft currents on our hands, then we have taken two soft limits, and in general the order of soft limits matters. It is instructive to consider the following limit of the celestial OPE of two outgoing positive-helicity gluons:
\be \ba & \lim_{\eps\to 0} \eta_1\eta_2\eps^2 \oh^{a,+}_{k+\eta_1\eps}(z_1, \zb_1) \oh^{b,+}_{\ell+\eta_2\eps}(z_2, \zb_2) \sim \\
& \frac{-i{f^{ab}}_c}{z_{12}} \hspace{-.7mm} \left[ \sum_{m=0}^{1-k} {2 \hspace{-.7mm} - \hspace{-.7mm} k \hspace{-.7mm} - \hspace{-.7mm} \ell \hspace{-.7mm} - \hspace{-.7mm} m \choose 1 \hspace{-.7mm} - \hspace{-.7mm} \ell} + \frac{(-)^{1-\ell}}{1 \hspace{-.7mm} + \hspace{-.7mm} \eta_2/\eta_1} \sum_{m=3-k-\ell}^\infty {k \hspace{-.7mm} - \hspace{-.7mm} 2 \hspace{-.7mm} + \hspace{-.7mm} m \choose 1 \hspace{-.7mm} - \hspace{-.7mm} \ell} \right] \frac{\zb_{12}^m}{m!} \p_{\zb_2}^m \oh^{c,+}_{(k+\ell-1)}(z_2, \zb_2). \ea \ee
One can recover the sequential soft limits by sending $\eta_1/\eta_2$ to zero or infinity, but leaving it general nicely displays how the limits fail to commute. The ``wedge" term, i.e. the sum on the right hand side with $0 \le m \le 1-k$, is independent of $\eta_1/\eta_2$, while the beyond-wedge sum with $m \ge 3-k-\ell$ has an overall factor depending on $\eta_1/\eta_2$. This in particular makes the holomorphic commutator of two soft gluons ill-defined, or rather it depends on how one takes the gluons soft.\footnote{Some authors use the simultaneous soft limit $\eta_1/\eta_2=1$ since it is the only pairwise-symmetric choice, but even so, one must still deal with the beyond-wedge term.} However note that the beyond-wedge sum involves $\p_{\zb_2}^m \oh^{c,+}_{(k+\ell-1)}(z_2, \zb_2)$ with $m \ge 3-k-\ell$. Recalling that all of $\oh^{c,+}_{(k+\ell-1)}$'s poles in $z_2$ are accompanied by a polynomial in $\zb_2$ of degree $1 - (k+\ell-1) = 2-k-\ell$, we see that $\p_{\zb_2}^m$ will kill these terms in the beyond-wedge sum, leaving only a polynomial in $z_2$. Therefore all subsequent holomorphic commutators of the beyond-wedge term will vanish. Similar comments apply for the holomorphic soft-soft OPE of gluons with arbitrary helicity. This allows us to salvage the holomorphic commutator of gluons in the following way \cite{Ball:2022bgg}. A single soft gluon has a well-defined action on hard insertions, and this defines a representation of the ostensible symmetry algebra. Write the action as
\be \oh^{a,s}_{(k)} \cdot \oh^{b,s'}_\Delta(z, \zb) \equiv \oint_z \frac{dw}{2\pi i} \oh^{a,s}_{(k)}(w, \wb) \oh^{b,s'}_\Delta(z, \zb). \ee
Acting repeatedly in this way on a hard insertion is also well-defined, and we can ask whether the action is equivariant with the holomorphic commutator of soft insertions. Explicitly, we ask about
\be \label{eq:equivar} \oh^{a_1, s_1}_{(k_1)} \cdot \left( \oh^{a_2,s_2}_{(k_2)} \cdot \oh^{b,s}_\Delta(z, \zb) \right) - \oh^{a_2,s_2}_{(k_2)} \cdot \left( \oh^{a_1,s_1}_{(k_1)} \cdot \oh^{b,s}_\Delta(z, \zb) \right) \stackrel{?}{=} [\oh^{a_1,s_1}_{(k_1)}, \oh^{a_2, s_2}_{(k_2)}] \cdot \oh^{b,s}_\Delta(z, \zb), \ee
where $[\oh^{a_1,s_1}_{(k_1)}, \oh^{a_2, s_2}_{(k_2)}]$ is the holomorphic commutator defined above. Note crucially that the right hand side is well-defined because the ambiguity in the holomorphic commutator drops out of the action on the hard insertion. If this equation is satisfied then the holomorphic commutator really is a commutator in the usual sense, modulo the beyond-wedge ambiguity discussed above. Consequently it must satisfy the Jacobi identity, and we can associate with it a symmetry algebra. One can show using the soft-hard OPEs above that in pure Yang-Mills this equation is indeed satisfied. Later on we will see an easier way to show this. In the positive-helicity sector one finds the current algebra \cite{Guevara:2021abz}
\be \ba [R^{k,a}_m, & R^{\ell,b}_n] \, \cdot \, \oh^{d,s}_\Delta(z, \zb) = \\
& -i f^{ab}{}_c {\frac{1-k}{2} - m + \frac{1-\ell}{2} - n \choose \frac{1-k}{2} - m} {\frac{1-k}{2} + m + \frac{1-\ell}{2} + n \choose \frac{1-k}{2} + m} R^{k+\ell-1,c}_{m+n}(z) \cdot \oh^{d,s}_\Delta(z, \zb). \ea \ee
We emphasize that there is no need to restrict attention to positive helicity. There is a consistent holomorphic current algebra involving gluons of both helicities. The non-commutativity between leading soft limits of gluons of opposite helicity is just a special case of the more general discussion above.

One can carry out similar steps for Einstein gravity to arrive at another consistent holomorphic current algebra. The positive-helicity sector can be massaged (or in more highbrow parlance, light transformed) to give a $w_{1+\infty}$-wedge current algebra \cite{Strominger:2021mtt}. The appearance of this algebra was especially interesting, because it already played an important role in many na\"ively unrelated contexts \cite{Pope:1989sr, Klebanov:1991hx}, in particular in twistor theory as the algebra of area-preserving diffeomorphisms on the Riemann sphere \cite{Penrose:1976js}. This connection has by now been thoroughly explored \cite{Adamo:2021lrv, Adamo:2021zpw, Monteiro:2022lwm, Bu:2022iak, Mason:2022hly}.

\section{Failing Jacobi}

Although we worked most explicitly with pure Yang-Mills, most of the steps discussed so far generalize straightforwardly to tree amplitudes in generic EFTs. A soft insertion can still be decomposed into holomorphic currents. However, it was shown in Refs. \cite{Mago:2021wje, Ren:2022sws} that the holomorphic commutator generically fails to obey \eqref{eq:equivar} and the Jacobi identity unless the EFT couplings satisfy certain algebraic relations. This initially came as a surprise, since usually in 2D CFT the standard contour pulling argument shows that the commutator of holomorphic charges must satisfy Jacobi. However, non-commutativity of soft limits obstructs this argument in the celestial context. In order to keep track of the order of soft limits, it will be useful to introduce the soft operator $S^{(k)}$, defined by
\be S^{(k)} \oh_\Delta(z, \zb) \equiv \oh_{(k)}(z, \zb). \ee
Now when taking multiple particles soft we can easily keep track of the order of limits by distinguishing e.g. $S_1^{(k)} S_2^{(\ell)} \oh_1 \oh_2$ versus $S_2^{(\ell)} S_1^{(k)} \oh_1 \oh_2$. One can show that in a general EFT the Jacobi identity for celestial soft currents can be boiled down to
\be \forall k,\ell \quad \res{z_2\to z_3} \res{z_1\to z_3} [S_1^{(k)}, S_2^{(\ell)}] \oh_1\oh_2\oh_3 \stackrel{?}{=} 0. \ee
In this form it is manifest that any violation of the Jacobi identity requires some non-commutativity of soft limits. But soft non-commutativity does not automatically violate Jacobi, as we have seen already for Yang-Mills. In more general EFTs the commutator of soft limits can have additional terms that spoil equivariance and the celestial Jacobi identity \cite{Ball:2022bgg}.

We will show this in the simple example of $\phi^3$ theory. A violation already occurs at the most leading order, which just involves the universal soft scalar theorem,
\be S_1^{(1)} \lang \oh_{\Delta_1} \oh_{\Delta_2} \dots \oh_{\Delta_n} \rang = \sum_{i=2}^n \frac{1}{z_{1i} \zb_{1i}} \lang \oh_{\Delta_2} \dots \oh_{\Delta_i-1} \dots \oh_{\Delta_n} \rang. \ee
We find
\be \ba \res{z_2\shortrightarrow z_3} \res{z_1\shortrightarrow z_3} [S_1^{(2)}, S_2^{(1)}] \, \oh_{\Delta_1} \oh_{\Delta_2} \oh_{\Delta_3} & = \res{z_2\shortrightarrow z_3} \res{z_1\shortrightarrow z_3} S_1^{(2)} \hspace{-1mm} \left( \hspace{-.2mm} \frac{1}{z_{12} \zb_{12}} \oh_{\Delta_1-1} \oh_{\Delta_3} \hspace{-1mm} + \hspace{-.5mm} \frac{1}{z_{23} \zb_{23}} \oh_{\Delta_1} \oh_{\Delta_3-1} \hspace{-.5mm} \right) \\
& = \res{z_2\shortrightarrow z_3} \res{z_1\shortrightarrow z_3} \frac{1}{z_{12} \zb_{12}} \, \frac{1}{z_{13} \zb_{13}} \oh_{\Delta_3-1} \\
& = \frac{-1}{\zb_{12} \zb_{13}} \oh_{\Delta_3-1} \ea \ee
where we used the facts that $S_1^{(2)} \oh_{\Delta_1} = 0$ and $S_1^{(2)} \oh_{\Delta_1-1} = S_1^{(1)} \oh_{\Delta_1}$. This non-vanishing case of $k=2, \, \ell=1$ shows that soft charges constructed in $\phi^3$ theory will not always satisfy the Jacobi identity. Other examples of EFTs failing Jacobi are Yang-Mills plus a $\phi F^2$ term with $\phi$ a real scalar, Yang-Mills plus an $F^3$ term, and Einstein gravity with analogous terms added \cite{Mago:2021wje}.

The failure of the Jacobi identity prevents us from associating soft currents with a symmetry algebra. One naturally wonders then what interpretation, if any, they admit. An enticing idea comes from work showing that the respective leading soft theorems in Yang-Mills, Einstein gravity, and the non-linear sigma model can be interpreted as generating parallel transport in the moduli space of bulk vacua (or dually, in the conformal manifold of celestial CFTs) \cite{Kapec:2022axw, Kapec:2022hih, Narayanan:2024qgb}. In these cases the commutator of soft limits is essentially the curvature on moduli space. However, upon closer inspection the connection to the present context seems less promising. The Jacobi identity is satisfied in Yang-Mills, Einstein gravity, and the non-linear sigma model, whereas our question is about interpreting soft currents in theories \it{failing} the Jacobi identity. Before doing so, we must introduce the double residue condition.

Our discussion so far has focused on \textit{soft} celestial amplitudes, where the holomorphic current and symmetry interpretations are most clear. But in practice it is usually more convenient to study the same questions, in particular the satisfaction of the Jacobi identity, at the level of hard celestial amplitudes or even hard momentum space amplitudes. It was argued in Ref. \cite{Ren:2022sws} and proved in Ref. \cite{Ball:2022bgg} that this can be achieved through the double residue condition. Explicitly, for tree-level amplitudes in a general EFT the following three conditions are equivalent:
\be \ba 1) \quad & \oh_{(k_1)} \cdot \left( \oh_{(k_2)} \cdot \oh_{\Delta_3}(z_3, \zb_3) \right) - \oh_{(k_2)} \cdot \left( \oh_{(k_1)} \cdot \oh_{\Delta_3}(z_3, \zb_3) \right) \stackrel{?}{=} [\oh_{(k_1)}, \oh_{(k_2)}] \cdot \oh_{\Delta_3}(z_3, \zb_3) \\
2) \quad & 0 \stackrel{?}{=} \left( \res{z_2\shortrightarrow z_3} \, \res{z_1\shortrightarrow z_2} - \res{z_1\shortrightarrow z_3} \, \res{z_2\shortrightarrow z_3} + \res{z_2\shortrightarrow z_3} \, \res{z_1\shortrightarrow z_3} \right) \oh_1(\omega_1, z_1, \zb_1) \oh_2(\omega_2, z_2, \zb_2) \oh_3(\omega_3, z_3, \zb_3) \\
3) \quad & 0 \stackrel{?}{=} \left( \res{z_2\shortrightarrow z_3} \, \res{z_1\shortrightarrow z_2} - \res{z_1\shortrightarrow z_3} \, \res{z_2\shortrightarrow z_3} + \res{z_2\shortrightarrow z_3} \, \res{z_1\shortrightarrow z_3} \right) \oh_{\Delta_1}(z_1, \zb_1) \oh_{\Delta_2}(z_2, \zb_2) \oh_{\Delta_3}(z_3, \zb_3). \ea \ee
The first statement here is the equivariance of soft currents discussed above, and it implies the Jacobi identity for the holomorphic commutator of soft currents. The second is the double residue condition on hard momentum space amplitudes, where by ``hard" we mean none of the $\omega_i$ have been taken soft. It is a statement about collinear splitting functions. The third statement is the double residue condition on hard celestial amplitudes, where by ``hard" we mean $\Delta_i \notin \integ$. It is a statement about the singular parts of celestial OPEs. In all cases we analytically continue to treat $z_i, \zb_i$ as independent variables. The equivalence is highly plausible on its face, and we leave its detailed derivation to the original reference \cite{Ball:2022bgg}. Our focus will be instead on the interpretation and consequences of the equivalence.

In all cases the conditions are naturally understood in terms of obstructions to the standard contour pulling argument for computing charge commutators from currents in 2D CFT. We have seen that for soft insertions the obstruction is the non-commutativity of soft limits, so that the currents are not actually simultaneously well-defined. We will see that for hard momentum space amplitudes the obstruction comes from the presence of extra poles due to three-particle factorization channels, and for hard celestial amplitudes the obstruction comes from the presence of branch cuts stretching between the $z_i$ \cite{Ball:2023sdz}.

\section{Double residue condition in momentum space}

The direct physical meaning of the double residue condition in momentum space is more obscure than the condition on soft insertions,\footnote{However, see Ref. \cite{Monteiro:2022lwm}.} but it is the easiest one to deal with in practice at the level of calculations. For example in pure Yang-Mills the full calculation for outgoing positive-helicity gluons is simply
\be \ba \res{z_2\shortrightarrow z_3} \, \res{z_1\shortrightarrow z_2} & \oh^{a,+}(\omega_1, z_1, \zb_1) \oh^{b,+}(\omega_2, z_2, \zb_2) \oh^{c,+}(\omega_3, z_3, \zb_3) \\
& = -i f^{ab}{}_d \frac{\omega_1 + \omega_2}{\omega_1\omega_2} \res{z_2\shortrightarrow z_3} \oh^{d,+}(\omega_1+\omega_2, z_2, \frac{\omega_1\zb_1 + \omega_2\zb_2}{\omega_1+\omega_2}) \oh^{c,+}(\omega_3, z_3, \zb_3) \\
& = -f^{ab}{}_d f^{dc}{}_e \frac{\omega_1+\omega_2}{\omega_1\omega_2} \frac{\omega_1+\omega_2+\omega_3}{(\omega_1+\omega_2)\omega_3} \oh^{e,+}(\omega_1+\omega_2+\omega_3, z_3, \frac{\omega_1\zb_1+\omega_2\zb_2+\omega_3\zb_3}{\omega_1+\omega_2+\omega_3}) \\
& = -f^{ab}{}_d f^{dc}{}_e \frac{\omega_1+\omega_2+\omega_3}{\omega_1 \omega_2 \omega_3} \oh^{e,+}(\omega_1+\omega_2+\omega_3, z_3, \frac{\omega_1\zb_1+\omega_2\zb_2+\omega_3\zb_3}{\omega_1+\omega_2+\omega_3}) \\
\res{z_1\shortrightarrow z_3} \, \res{z_2\shortrightarrow z_3} & \oh^{a,+}(\omega_1, z_1, \zb_1) \oh^{b,+}(\omega_2, z_2, \zb_2) \oh^{c,+}(\omega_3, z_3, \zb_3) \\
& = -f^{bc}{}_d f^{ad}{}_e \frac{\omega_1+\omega_2+\omega_3}{\omega_1 \omega_2 \omega_3} \oh^{e,+}(\omega_1+\omega_2+\omega_3, z_3, \frac{\omega_1\zb_1+\omega_2\zb_2+\omega_3\zb_3}{\omega_1+\omega_2+\omega_3}) \\
\res{z_2\shortrightarrow z_3} \, \res{z_1\shortrightarrow z_3} & \oh^{a,+}(\omega_1, z_1, \zb_1) \oh^{b,+}(\omega_2, z_2, \zb_2) \oh^{c,+}(\omega_3, z_3, \zb_3) \\
& = -f^{ac}{}_d f^{bd}{}_e \frac{\omega_1+\omega_2+\omega_3}{\omega_1 \omega_2 \omega_3} \oh^{e,+}(\omega_1+\omega_2+\omega_3, z_3, \frac{\omega_1\zb_1+\omega_2\zb_2+\omega_3\zb_3}{\omega_1+\omega_2+\omega_3}). \ea \ee
We see that the double residue condition reduces to the Jacobi identity on the structure constants themselves. There is a similar cancellation for any combination of helicities.

\begin{figure}
  \centering
  \begin{minipage}[b]{0.31\textwidth}
    \centering
        \includegraphics[width=\textwidth]{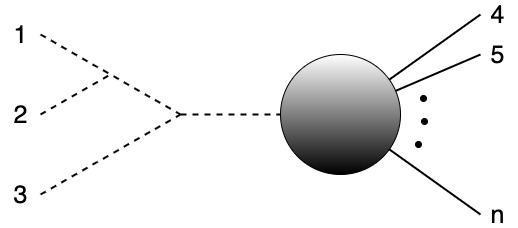}
  \end{minipage}
  \hfill
  \begin{minipage}[b]{0.31\textwidth}
  \centering
        \includegraphics[width=\textwidth]{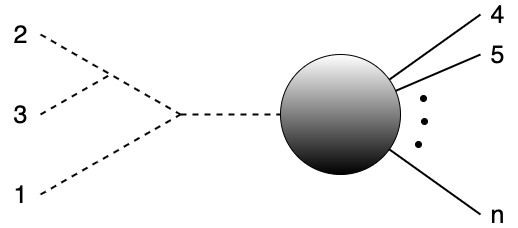}
  \end{minipage}
  \hfill
  \begin{minipage}[b]{0.31\textwidth}
  \centering
        \includegraphics[width=\textwidth]{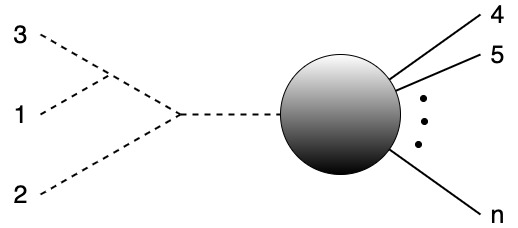}
  \end{minipage} 
  \caption{\label{fig:feyndias} Feynman diagrams that can contribute to the double residue condition on particles $1, 2, 3$. Dotted lines indicate massless bosons, solid lines indicate any particle in the theory, and the shaded spheres indicate any tree-level process.}
\end{figure}

\begin{figure}
\includegraphics[width=\textwidth]{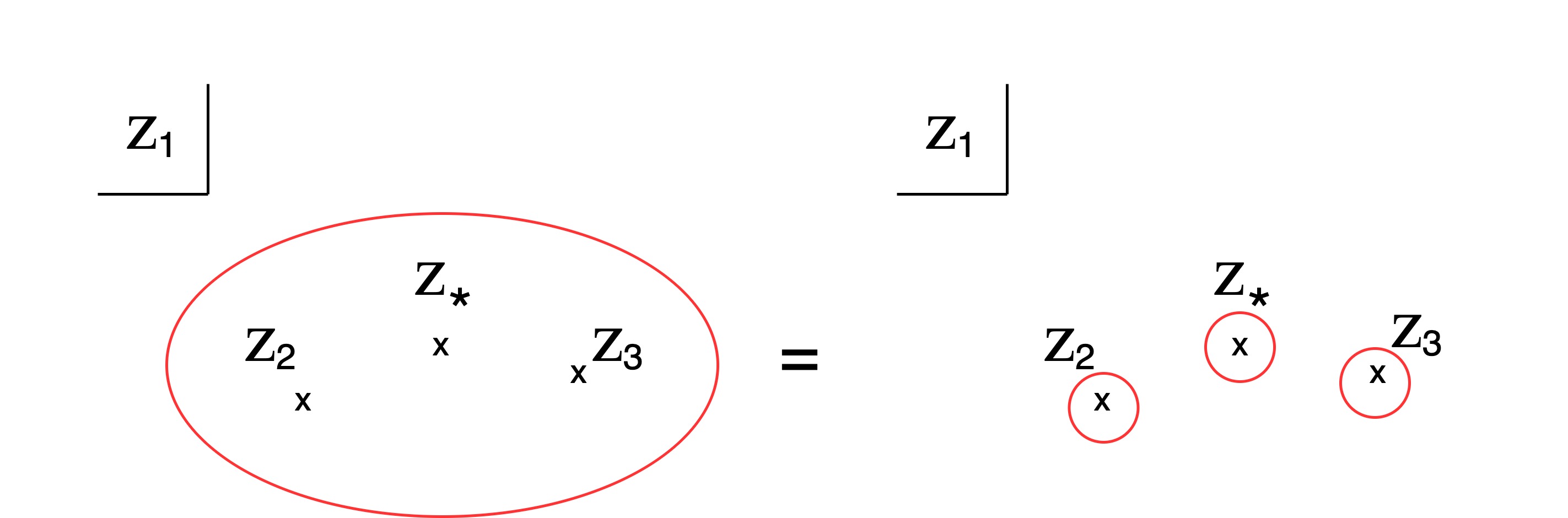}
\caption{\label{fig:pole_pull} Equivalent contours around poles in the complex $z_1$ plane are shown. The crux of the usual contour pulling argument is that a single contour encircling poles at $z_2$ and $z_3$ is equivalent to two small contours separately encircling $z_2$ and $z_3$. Here the presence of the pole at $z_*$ modifies this relation.}
\end{figure}

In a general theory one can always follow one's nose and evaluate the double residue condition in terms of collinear splitting functions, but understanding the condition in terms of the contour pulling argument provides further insight. Towards this, choose any three massless bosons and view the amplitude as a rational function of their parameters $z_1$, $z_2$, and $z_3$. Evaluating the double residues only requires knowledge of the amplitude in a neighborhood of $z_1=z_2=z_3$. In this neighborhood, generically the only poles are the collinear ones where one of $z_{12}$, $z_{13}$, or $z_{23}$ vanishes and the three-particle factorization one at $(p_1+p_2+p_3)^2 = 0$. A given diagram can only contain one of the collinear poles, so the second residue must come from three-particle factorization. Therefore only the diagrams in Fig. \ref{fig:feyndias} contribute to the double residue condition.
It is instructive to write out
\be \frac{1}{(p_1+p_2+p_3)^2} = \frac{-1/4}{\epsilon_1 \epsilon_2 \omega_1\omega_2z_{12}\zb_{12} + \epsilon_1 \epsilon_3 \omega_1\omega_3z_{13}\zb_{13} + \epsilon_2 \epsilon_3 \omega_2\omega_3z_{23}\zb_{23}}. \ee
We see it gives a pole at $z_1 = z_*$ for some $z_*$ that we will not bother to write out explicitly. The point is that this pole is ``nonlocal" in the sense that $z_* \ne z_i$ for any of the other particles. Since we are not in a conformal basis, this sense of nonlocality does not necessarily indicate any inconsistency in the theory. We will return to this point in the next section. But the nonlocality of the $z_1 = z_*$ pole \it{can} obstruct the contour pulling argument, as shown in Fig. \ref{fig:pole_pull}.
We emphasize however that the mere existence of a pole at $z_*$, i.e. a three-particle factorization channel, does not automatically imply the failure of the double residue condition. There can be cancellations, as we have already seen for Yang-Mills.

To understand when exactly such cancellations occur, it helps to start by rewriting the double residue condition as\footnote{This can be verified explicitly by writing out the analytic structure of a general diagram \cite{Ball:2022bgg}.}
\be \ba & -\res{z_2 \to z_3} \res{z_1 \to z_*} \oh_1(\omega_1, z_1, \zb_1) \oh_2(\omega_2, z_2, \zb_2) \oh_3(\omega_3, z_3, \zb_3) \\
& \,\,\, = \left( \res{z_2\shortrightarrow z_3} \, \res{z_1\shortrightarrow z_2} - \res{z_1\shortrightarrow z_3} \, \res{z_2\shortrightarrow z_3} + \res{z_2\shortrightarrow z_3} \, \res{z_1\shortrightarrow z_3} \right) \oh_1(\omega_1, z_1, \zb_1) \oh_2(\omega_2, z_2, \zb_2) \oh_3(\omega_3, z_3, \zb_3). \ea \ee
The residue $\res{z_1\to z_*}$ simply grabs from the $(p_1+p_2+p_3)^2=0$ factorization channel, giving
\be \res{z_1\to z_*} \ca_{123\dots} = \delta^{(4)}\left( \sum_{i=1}^n p_i \right) \sum_I \hat\ca_{123I} \frac{-1/4}{\epsilon_1\omega_1 (\epsilon_2\omega_2\zb_{12}+\epsilon_3\omega_3\zb_{13})} \hat\ca_{-I\dots} \ee
where the sum is over intermediate particles and helicities. In the latter amplitude $-I$ indicates the CPT conjugate of particle $I$, with momentum $-p_I$. The subsequent residue on $z_{23}$ comes only from $\hat\ca_{123I}$. It turns out that the entire double residue condition boils down to this single residue on four-point amplitudes $\hat\ca_{123I}$, and further that if $\hat\ca_{123I}$ is written with spinor-helicity variables then the residue is the part whose angle bracket weight equals $-1$, where angle bracket weight is defined as the number of angle brackets products in the numerator, minus the number in the denominator. For example in Yang-Mills a color-ordered four-point MHV amplitude can be written (up to coupling constants)
\be \hat\ca_{++--} = \frac{\lang 34 \rang^3}{\lang 12 \rang \lang 23 \rang \lang 41 \rang}, \ee
and it has angle bracket weight $3 - 3 = 0$. Angle bracket weight is unchanged by using momentum conservation to rewrite the amplitude, so it is well-defined. The punchline is that the double residue condition on bosons $1, 2, 3$ is satisfied if and only if the angle bracket weight $-1$ part of $\hat\ca_{123I}$ vanishes for all particles $I$. We can use this newfound tool to very quickly show that the double residue condition on positive-helicity gluons in Yang-Mills is satisfied. The only four-point amplitudes to check are $\hat\ca_{+++\pm}$, and they both vanish identically, so the double residue condition must be satisfied. One can also show easily that the double residue condition is satisfied for gluons of any helicities.

\section{Double residue condition for hard celestial amplitudes}

\begin{figure}
\includegraphics[width=\textwidth]{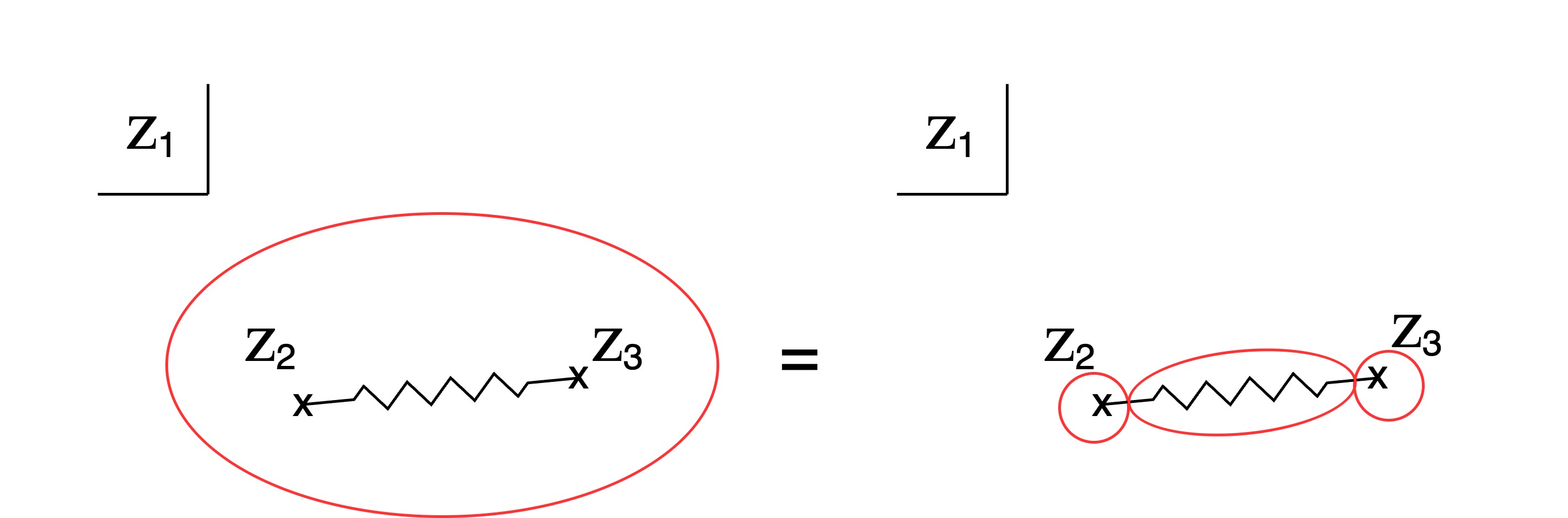}
\caption{\label{fig:branch_pull} Equivalent contours around singularities and a branch cut in the complex $z_1$ plane are shown. The contour encircling the singularities at $z_2$ and $z_3$ also includes the branch cut stretching between them, and it contributes to the contour integral.}
\end{figure}

Once again it is instructive to ask ourselves how the contour pulling argument can fail. Unlike in the soft case, we know that celestial correlators of hard insertions are well-defined. And unlike in the momentum space case, we expect the $z$-dependence to have singularities only at the locations of other insertions.\footnote{The intuition here is that any poles whose location depends on the $\omega_i$ will be smeared by the Mellin integrals, turning them into less singular objects like branch cuts. A toy version of this is the integral $\int_0^1 \frac{d\omega}{\omega - z} = \log\frac{z-1}{z}$. The integrand's pole at $z=\omega$ smooths out to a branch cut stretching from $z=0$ to $z=1$.} This seemingly leaves only one possible obstruction to the contour pulling argument: branch cuts stretching between insertions. The violation of the double residue condition is then essentially the integrated discontinuity of the branch cut. See Fig. \ref{fig:branch_pull}.
The point is that when a branch cut is present, this will generically be nonzero.

The presence of branch cuts in theories failing the double residue condition was confirmed in Ref. \cite{Ball:2023sdz}. Working with full celestial amplitudes is rather intractable, which is why the usual (holomorphic) celestial OPE is so useful. It contains a wealth of information while also being computationally convenient. However, it is not quite powerful enough to directly see the branch cuts of current interest. The authors of Ref. \cite{Ball:2023sdz} leveraged the fact that evaluating the double residue condition requires only knowledge of the leading singularity in a neighborhood of the point $z_1=z_2=z_3$ in parameter space. They formalized the notion of zooming in on this point by defining the holomorphic multicollinear limit. The usual holomorphic collinear limit zooms in on the point $z_1=z_2$, and it has a straightforward generalization to any number of coincident $z_i$. There is a hierarchy of information, with the $3$-collinear limit knowing more than consecutive $2$-collinear limits, and so on. The authors also worked out the celestial dual of the holomorphic multicollinear limit, namely the holomorphic celestial multi-OPE. They found that when the double residue condition is satisfied, the 3-OPE's dependence on $z_1$, $z_2$, $z_3$ is single-valued. Conversely, in theories failing the double residue condition the 3-OPE acquires a branch cut from the Mellin transform of the three-particle propagator $\frac{1}{(p_1+p_2+p_3)^2}$. This was argued in general and confirmed explicitly in $\phi^3$ theory.

\subsection{New term in celestial 2-OPE} \label{sec:newOPE}

The presence of branch cuts in the 3-OPE is in tension with their absence in the usual 2-OPE \cite{Fan:2019emx, Pate:2019lpp, Himwich:2021dau}. The resolution is that the 2-OPE in Ref. \cite{Himwich:2021dau} is actually missing a term with a branch cut. This indicates new primary content in the fusion of two celestial insertions, and the authors of Ref. \cite{Ball:2023sdz} proposed that it corresponds to a multi-particle insertion. This was further studied in Ref. \cite{Guevara:2024ixn}. We now explain in detail how this new OPE term arises, and how it was originally missed. Consider a generic celestial amplitude
\be \tilde \ca_n = \int_0^\infty d\omega_n \omega_n^{\Delta_n-1} \dots \int_0^\infty d\omega_2 \omega_2^{\Delta_2-1} \int_0^\infty d\omega_1 \omega_1^{\Delta_1-1} \ca_n \ee
and study it at small $z_{12}$. The integrand, i.e. the amplitude $\ca_n$, has a two-particle factorization channel when $(p_1 + p_2)^2 = -4\omega_1\omega_2 z_{12} \zb_{12}$ is small. The leading term in this expansion is the collinear splitting function, and its Mellin transform gives the standard celestial OPE. But note that no matter how small $z_{12}$ is, if $\omega_1\omega_2$ is large enough then $(p_1+p_2)^2$ is not small. Since the integration range on $\omega_1, \omega_2$ goes to infinity, there will \it{always} be a region where $(p_1+p_2)^2$ is not small, and therefore we cannot assume that the collinear splitting function is the dominant contribution to the integral. An instructive toy example of this phenomenon comes from the integral representation of the ${}_2F_1$ hypergeometric function,
\be {}_2F_1(a, b; c; z) = \frac{1}{B(b, c-b)} \int_0^1 w^{b-1} (1-w)^{c-b-1} (1 - zw)^{-a} dw. \ee
The function is analytic near $z=0$, but near $z=1$ it has the more exotic behavior
\be \label{eq:2F1z1} \ba {}_2F_1&(a, b; c; z) = \\
& \left( \frac{B(b,c-b-a)}{B(b,c-b)} + O(1-z) \right) + (1-z)^{c-b-a} \left( \frac{B(c-b,a+b-c)}{B(b,c-b)} + O(1-z) \right). \ea \ee
Note the latter term has a branch cut when $c-b-a$ is non-integer. Also, the two terms exchange dominance depending on the sign of the real part of $c-b-a$. Both of these terms can be deduced from the integral representation, but one must be careful. The integrand is analytic in a neighborhood of $z=1$ as long as $w<1$, so it is equal to its Taylor series in $1-z$ and we have
\be {}_2F_1(a, b; c; z) = \frac{1}{B(b,c-b)} \int_0^1 dw \Big( w^{b-1} (1-w)^{c-b-a-1} + O(1-z) \Big). \ee
The key point is that we cannot quite move the integral inside the sum. If we could, we would conclude that the ${}_2F_1$ is analytic near $z=1$ with leading behavior
\be {}_2F_1(a, b; c; z) \stackrel{?}{=} \frac{B(b,c-b-a)}{B(b,c-b)} + O(1-z). \ee
This correctly obtains the first term in \eqref{eq:2F1z1}, but it entirely misses the second term. This is perfectly analogous to how the original derivations of the celestial OPE used the collinear expansion on the integrand and then moved it past the Mellin integral \cite{Fan:2019emx, Pate:2019lpp, Himwich:2021dau}. In both cases the red flag is that the ``leading" term in the expansion of the integrand is not actually dominant on the entire integration range. Here this is because $(1-zw)^{-a}$ has a branch point at $z=1/w$, and as $w\to 1$ this branch point approaches $z=1$. Before moving on, we note that the integral giving $\frac{B(b,c-b-a)}{B(b,c-b)}$ converges only for $c-b-a>0$,\footnote{Convergence also requires $b>0$.} which happens to be the range where the first term in \eqref{eq:2F1z1} does indeed dominate the second term.

We can obtain the second term in \eqref{eq:2F1z1} as follows. Change the integration variable from $w$ to $\sig$ via $w = 1 - (1-z)\sig$. For the moment assume $z$ is real and $z<1$; later we can analytically continue to arbitrary values. With this substitution the integral becomes
\be \ba {}_2F_1(a, b; c; z) 
& = \frac{(1-z)^{c-b-a}}{B(b, c-b)} \int_0^{\frac{1}{1-z}} (1-(1-z)\sig)^{b-1} \sig^{c-b-1} (1+z\sig)^{-a} d\sig. \ea \ee
To get the leading contribution as $z\to 1$ we simply set $z=1$ in the integral, yielding
\be \frac{(1-z)^{c-b-a}}{B(b, c-b)} \int_0^\infty \sig^{c-b-1} (1+\sig)^{-a} d\sig = \frac{(1-z)^{c-b-a}}{B(b, c-b)} B(c-b, a+b-c). \ee
This recovers the second term in \eqref{eq:2F1z1}. Note convergence of the integral here requires $c-b-a < 0$,\footnote{Convergence also requires $c-b>0$.} which is indeed the range where the second term in \eqref{eq:2F1z1} dominates. Overall we have learned that Taylor expanding the integrand is a legitimate way to find the leading behavior of the integral in a parameter range where it converges, but that this method misses important subleading terms that may become dominant in other parameter ranges. Along these lines it is useful to distinguish between the collinear part of a celestial (multi-)OPE versus the full (multi-)OPE. The familiar celestial OPE of Ref. \cite{Himwich:2021dau} is really only the collinear part. Techniques similar to those shown here were used to uncover a branch cut term in the 3-collinear part of the celestial 3-OPE of $\phi^3$ theory, indicating in turn that the full 2-OPE must have a branch cut. In general branch cuts are expected in the celestial OPEs of all theories failing the double residue condition \cite{Ball:2023sdz}. We emphasize however that failure of the double residue condition is not an indication of any pathology. It has a more benign interpretation, to which we now turn.

\subsection{Hard (celestial) currents}

The double residue condition for hard celestial amplitudes is closely related to the question of to what extent a hard celestial insertion constitutes a holomorphic current. The first hint that hard insertions might behave as currents comes from the form of the collinear part of the celestial OPE, which always comes with a $1/z$ pole, resembling a current algebra. But the core idea behind hard currents is actually very general, and we can discuss it in the context of a generic 2D CFT, celestial or not. In any 2D CFT one can take a local operator $\oh(z, \zb)$ and analytically continue to view it as a function of $z$ with $\zb$ as a fixed parameter. The resulting $z$-dependence will be locally holomorphic and it will have singularities only at the locations of other insertions. This sounds very much like the behavior of a holomorphic current. We discussed above how celestial soft insertions can be viewed as $\zb$-weighted sums of holomorphic currents, but there the $\zb$-dependence was polynomial. In this more general context the $\zb$-dependence can take almost any form, and so the term ``hard current" should be taken with a sizable pinch of salt. But still, hard currents can give rise to consistent charge algebras and reveal genuine symmetries. The only way for this construction to fail is if the analytically continued $z$-dependence is not single-valued, i.e. if it has branch cuts.\footnote{Before complexification, i.e. when $\zb = z^*$, all CFT correlators are single-valued. Any branch cuts in $z$ are cancelled by branch cuts in $\zb$. It is only after analytic continuation that correlators can fail to be single-valued.} The presence of branch cuts is entirely plausible, or even expected, when the dimension of the operator is non-integer. For example consider the 2D critical Ising model. Its three primaries are the identity $\textbf{1}$ of weights $(0, 0)$, the spin operator $\sig$ of weights $(\frac{1}{16}, \frac{1}{16})$, and the energy operator $\eps$ of weights $(\half, \half)$. The spin-spin two-point function is
\be \lang \sig(z, \zb) \sig(0, 0) \rang = \frac{1}{(z \zb)^{1/8}} \ee
and we see that its $z$-dependence of $z^{-1/8}$ has a branch cut stretching from $z=0$ to $z=\infty$. Similarly in a general correlator we expect $\sig$'s $z$-dependence to have branch cuts, and so contour integrals like $\oint \frac{dz}{2\pi i} \, z^m \sig(z, \zb)$ will be ill-defined and will not provide conserved charges.\footnote{For any contour of fixed $|z|$ one can choose $m \in \real$ appropriately so that $z^m \sig(z, \zb)$ is single-valued, but the point is that no fixed value for $m$ will work for all $|z|$ simultaneously in general. Then objects like commutators of charges are ill-defined.} What about the energy operator $\eps$? Its two-point function,
\be \lang \eps(z, \zb) \eps(0, 0) \rang = \frac{1}{z \zb}, \ee
actually has single-valued $z$-dependence. This is encouraging, and raises the question of whether $\eps(z, \zb)$ can be interpreted as a $\zb$-weighted sum of holomorphic currents. Our hopes seem to be dashed on the three-point function
\be \lang \sig(z_1, \zb_1) \sig(z_2, \zb_2) \eps(z_3, \zb_3) \rang = \half |z_{12}|^{3/4} |z_{13}|^{-1} |z_{23}|^{-1}, \ee
whose $z_3$-dependence has a branch cut stretching between $z_1$ and $z_2$. However, if we restrict to the sector of the theory including only $\textbf{1}, \eps$ insertions and no $\sig$ insertions, then $\eps$'s $z$-dependence remains single-valued. In fact the fusion rule $\eps \times \eps = \textbf{1}$ implies the slightly stronger statement that in a correlator $\lang \eps(z, \zb) \dots \rang$ including arbitrary other insertions, the $z$-dependence will be single-valued around all other $\eps$ insertions so that branch cuts only stretch between $\sig$ insertions. With this restriction to action on $\textbf{1}, \eps$ insertions we can proceed with our attempt to treat $\eps$ as a holomorphic current. The ``charges"
\be Q_m(\zb) \equiv \oint_{\mc C} \frac{dz}{2\pi i} z^{m-\half} \eps(z, \zb), \qquad m \in \integ + \half \ee
are well-defined for $m\in\integ+\half$ as long as the contour $\mc{C}$ only encloses insertions of $\textbf{1}, \eps$, and their commutator satisfies the Jacobi identity. It is instructive to consider their action on the field $\eps(z', \zb')$,
\be [Q_m(\zb), \eps(z', \zb')] = \oint_{z'} \frac{dz}{2\pi i} z^{m-\half} \eps(z, \zb) \eps(z', \zb'). \ee
For concreteness let us study this using the four-point correlator
\be \lang \eps(z_1, \zb_1) \eps(z_2, \zb_2) \sig(z_3, \zb_3) \sig(z_4, \zb_4) \rang = \left| z_{12}^{-1} z_{34}^{-1/8} (z_{13} z_{24} z_{14} z_{23})^{-1/2} (z_{13} z_{24} - \half z_{12} z_{34}) \right|^2 \ee
with $|z_1|, |z_2| < |z_3|, |z_4|$. We have
\be \ba \oint_{z_2} \frac{dz_1}{2\pi i} z_1^{m-\half} \lang & \eps(z_1, \zb_1) \eps(z_2, \zb_2) \sig(z_3, \zb_3) \sig(z_4, \zb_4) \rang = \\
& \left( z_2^{m-\half} z_{34}^{-1/8} \right) \left( \zb_{12}^{-1} \zb_{34}^{-1/8} (\zb_{13} \zb_{24} \zb_{14} \zb_{23})^{-1/2} (\zb_{13} \zb_{24} - \half \zb_{12} \zb_{34}) \right). \ea \ee
We see that the anti-holomorphic part of $\lang \eps_1\eps_2\sig_3\sig_4 \rang$ was unchanged, while the holomorphic part turned into $z_2^{m-\half}$ times the holomorphic part of $\lang \sig_3\sig_4 \rang = |z_{34}|^{-1/4}$. This suggests that in some sense the action of $Q_m(\zb)$ on $\eps$ is a symmetry implementing a holomorphic shift of $\eps$ by $z^{m-\half}$. One could be forgiven for initially finding this somewhat speculative, but we will make it completely precise using the formulation of the 2D Ising CFT in terms of free fermions $\psi(z)$ and $\tilde\psi(\zb)$. Most properties of the free fermion can be deduced from the action
\be S_{\rm ferm} = \frac{1}{4\pi} \int d^2z (\psi \bar\p \psi + \tilde\psi \p \tilde\psi). \ee
We of course have the fermion fields $\psi(z)$ and $\tilde\psi(\zb)$, respectively of weights $(\half, 0)$ and $(0, \half)$. We also have the identity operator $\textbf{1}$, which corresponds to the Neveu-Schwarz vacuum. Finally we also have the spin fields $S(z)$ and $\tilde S(\zb)$, respectively of weights $(\frac{1}{16}, 0)$ and $(0, \frac{1}{16})$, which correspond to the Ramond vacua. The non-identity operators here are not truly local operators since they are not single-valued in general. However certain combinations of them \it{are} single-valued, and give precisely the primary content of the 2D Ising CFT. We can identify
\be \ba \textbf{1} \quad & \longleftrightarrow \quad \textbf{1}, \\
\sig(z, \zb) \quad & \longleftrightarrow \quad S(z) \tilde S(\zb), \\
\eps(z, \zb) \quad & \longleftrightarrow \quad \psi(z) \tilde\psi(\zb). \ea \ee
From this perspective the meaning of analytically continuing $\eps(z, \zb)$ is clear: it just amounts to two insertions $\psi(z)$ and $\tilde\psi(\zb)$ that are no longer at the same point. In the sector including only $\bf{1}, \eps$ insertions there is no branch cut connecting them. So in this sector $\eps(z, \zb)$ really does include a holomorphic current $\psi(z)$, as well as the anti-holomorphic factor $\tilde\psi(\zb)$ that goes along for the ride. We deduced above that the effect of our charge on the holomorphic part of $\eps(z, \zb)$ was to shift it by $z^{m-\half}$. This apparently corresponds to $\psi(z) \to \psi(z) + z^{m-\half}$, and such a holomorphic shift is manifestly a symmetry of the action $S_{\rm ferm}$. So the admittedly indirect approach of constructing charges from hard currents did unveil a genuine symmetry in the end.

To summarize what we have learned, any operator in 2D CFT can be analytically continued in $z$ while viewing $\zb$ as a fixed parameter, and the only obstruction to interpreting it as a (very generalized) holomorphic current is the possibility of branch cuts. The 2D Ising CFT has two nontrivial primaries. The spin operator $\sig$ cannot be interpreted as a current due to its branch cuts, and therefore our methods find no symmetry associated with it. This of course does not mean that the 2D Ising CFT is problematic in any way. Conversely, the energy operator $\eps$ has no branch cuts around $\bf{1}, \eps$ insertions and this allows us to interpret it as a holomorphic current with an associated symmetry.

These lessons apply directly to celestial CFT. Hard insertions failing the double residue condition have branch cuts preventing the construction of holomorphic charges from them, and consequently they have no symmetry interpretation. Note that branch cuts still cause such obstructions even when they are small enough that the ``residue" is unaffected. For example consider the function $\frac{1}{z_{12}^2} \log \frac{z_{13}}{z_{23}}$. It has a branch cut in $z_1$ stretching from $z_3$ to $\infty$, yet small contour integrals in $z_1$ around $z_2$ and $z_3$ are still well-defined. Nevertheless it fails the double residue condition,
\be \left( \res{z_2\shortrightarrow z_3} \, \res{z_1\shortrightarrow z_2} - \res{z_1\shortrightarrow z_3} \, \res{z_2\shortrightarrow z_3} + \res{z_2\shortrightarrow z_3} \, \res{z_1\shortrightarrow z_3} \right) \frac{1}{z_{12}^2} \log \frac{z_{13}}{z_{23}} = 1 - 0 + 0. \ee
On the other hand, hard insertions satisfying the double residue condition, even if only in a subsector of the theory, have no branch cuts in their 3-OPEs and consequently \textit{can} be interpreted as generalized holomorphic currents. The current algebra data are simply the standard celestial OPE coefficients. For example in Einstein gravity the celestial OPE of two outgoing positive-helicity gravitons is
\be G^+_{\Delta_1}(z_1, \zb_1) G^+_{\Delta_2}(z_2, \zb_2) \sim \frac{1}{z_{12}} \sum_{m=0}^\infty B(\Delta_1 - 1 + m, \Delta_2 - 1) \frac{\zb_{12}^{m+1}}{m!} \p_{\zb_2}^m G^+_{\Delta_1+\Delta_2}(z_2, \zb_2) \ee
and we interpret the entire coefficient multiplying $\frac{1}{z_{12}}$ as the current algebra data. This offers a generalization of the paradigm in Ref. \cite{Guevara:2021abz}. However it may be only a modest generalization since the soft expansion contains the same information as the original amplitude.

Let us finally return to the question of how to interpret celestial soft currents failing the Jacobi identity. We have seen that the Jacobi identity for soft currents is equivalent to the double residue condition for hard currents, whose failure precludes a symmetry algebra interpretation for the hard currents but has no ill consequences for the OPE. Therefore our stance is that failure of the Jacobi identity by a celestial CFT is not a problem of the theory, but simply a property of it. Admittedly it still seems somewhat strange that a single soft insertion resembles a (sum of) holomorphic currents yet may not have a symmetry interpretation, but we note that in a certain sense a lone holomorphic current is cheap and unremarkable. For example in the 2D Ising CFT the spin operator's two-point function $\lang \sig(z, \zb) \sig(0, 0) \rang = (z \zb)^{-1/8}$ has a branch cut in $z$, but we could still define conserved charges $\oint_0 \frac{dz}{2\pi i} z^{m+1/8} \sig(z, \zb)$ acting on $\sig(0, 0)$, where $m\in\integ$. We do not normally attempt to associate these with any symmetry, so it seems plausible to us that a single soft insertion in a celestial CFT failing the Jacobi identity is also not associated with any symmetry.

\subsection{Two notions of associativity}

A common misconception is that associativity of the OPE is the same as associativity of the action of charges.\footnote{Recall that any commutator arising from an underlying associative multiplication automatically satisfies the Jacobi identity.} Here we mean OPE associativity in the bootstrap sense, that applying the OPE in different orders must give the same result. This involves the whole infinite sum over the OPE's regular terms. In contrast, computing the action of a charge requires only the singular part of the OPE.

Of course, usual conserved charges in CFT \it{do} act associatively and thus their commutator must satisfy the Jacobi identity. As we have seen, the real question is not whether charges act associatively, but whether the celestial charges even (co)exist. The example of 2D Ising showed that hard currents do make sense when there are no branch cuts in the neighborhood of interest, but when branch cuts are present the na\"ive charge construction may not even exist. Despite this, the OPE of 2D Ising is certainly associative in the bootstrap sense.

It is also worth noting here that there are two unrelated notions of ``crossing symmetry", one being equivalent to the bootstrap sense of OPE associativity and the other referring to the invariance of amplitudes upon swapping an incoming particle for its outgoing CPT conjugate with opposite four-momentum.\footnote{See, however, Ref. \cite{Caron-Huot:2023ikn} for subtleties and caveats in crossing symmetry of amplitudes.} Since celestial CFT involves both amplitudes and CFT, confusion sometimes arises over this terminology.

\section{Conclusion}

In this work we have reviewed currents in celestial CFT, with an emphasis on the consistency of their symmetry interpretation. After discussing the extent to which soft insertions constitute (sums of) holomorphic currents we studied the Jacobi identity of their charges, which somewhat counterintuitively can fail. The Jacobi identity on soft currents is equivalent to the double residue condition on hard momentum space and celestial amplitudes, and its failure was understood in terms of obstructions to the contour pulling argument. In momentum space the obstruction came from multi-particle factorization poles, while for hard celestial amplitudes these poles transformed into branch cuts. In the soft case non-commutativity of soft limits was the culprit, which itself can be traced to the poles/branch cuts in the hard cases. Most theories satisfying the Jacobi identity, such as Yang-Mills and Einstein gravity, still have some non-commutativity of soft limits, but in generic theories the non-commutativity is even more violent and leads to failure of the Jacobi identity.

We worked through explicit examples, new to the literature, illustrating how branch cuts missed in Ref. \cite{Himwich:2021dau} can arise and introducing the notion of hard currents. We showed using the integral representation of the ${}_2F_1$ hypergeomtric function that Taylor expanding the integrand can miss terms including branch cuts, and we explained how this same concept applies to the celestial OPE and its non-collinear parts. We also showed in the 2D Ising CFT that the energy operator $\eps$ makes sense as a hard current and genuinely reveals an underlying symmetry, while the spin operator $\sig$ has no current interpretation. Carrying these lessons to celestial CFT, we concluded that failure of the Jacobi identity simply indicates the lack of a symmetry, and is not necessarily a problem for the OPE or the theory generally.

Throughout this work we focused on the special case of massless amplitudes at tree level. The massive and loop level cases are important, but are much less understood. For recent progress on the massive case see Ref. \cite{Himwich:2023njb}. Interestingly the authors found branch cuts in the analytically continued celestial amplitudes that may be related to those discussed above. For progress at one loop see Refs. \cite{Ball:2021tmb, Bhardwaj:2022anh, Bittleston:2022jeq, Ball:2023qim, Bhardwaj:2024wld}. There are also various twistor constructions for soft algebras that hold at all loop orders, often closely related to self-dual theories \cite{Costello:2022upu, Bu:2022dis, Costello:2022jpg, Bittleston:2023bzp, Costello:2023hmi, Bu:2023vjt}.

\section*{Acknowledgments}

We are grateful to Sabrina Pasterski for comments on the draft, and to Modern Physics Letters A for the invitation to review this topic. The author is supported by the Celestial Holography Initiative at the Perimeter Institute for Theoretical Physics and the Simons Collaboration on Celestial Holography. Research at the Perimeter Institute is supported by the Government of Canada through the Department of Innovation, Science and Industry Canada and by the Province of Ontario through the Ministry of Colleges and Universities.

\bibliography{cel-rev}
\bibliographystyle{utphys}

\end{document}